\documentclass[aps,prl,twocolumn,showpacs,amsmath,amssymb]{revtex4}
\usepackage{amssymb}
\usepackage{graphicx}
\usepackage{dcolumn}
\usepackage{verbatim}
\begin{document}

\title{Molecular-spin dynamics study of electromagnons in
multiferroic RMn$_2$O$_5$}

\author{Kun Cao}
\affiliation{Key Laboratory of Quantum Information, University of
Science and Technology of China, Hefei, 230026, Anhui, People's Republic of
China}

\author{Guang-Can Guo}
\affiliation{Key Laboratory of Quantum Information, University of
Science and Technology of China, Hefei, 230026, Anhui, People's Republic of
China}

\author{Lixin He}
\email{helx@ustc.edu.cn}
\affiliation{Key Laboratory of Quantum Information, University of
Science and Technology of China, Hefei, 230026, Anhui, People's Republic of
China}

\date{\today}

\begin{abstract}
We investigate the electromagnon in magnetoferroelectrics RMn$_2$O$_5$ using
combined molecular-spin dynamics simulations. We identify the origin of the
electromagnon modes observed in the optical spectra and reproduce
the most salient features of the electromagnon in these compounds.
We find that the electromagnon frequencies are very sensitive to the
magnetic wave vector along the $a$ direction. We further investigate the
electromagnon in magnetic field. Although the modes frequencies change
significant under magnetic field, the static dielectric constant
electromagnon does not change much in the magnetic field.
\end{abstract}

\pacs{75.85.+t, 77.80.-e,  63.20.-e}
\maketitle

Magnetoferrelectrics with strong magnetoelectric (ME) couplings have
attracted intensive attention for their novel physics
\cite{hur04b,kimura03,goto04,hur04} and potential applications in
tunable multifunctional devices\cite{cheong07,fiebig05}. One of the most
attractive features of the magnetoferroelectrics is that
the spin waves and infrared phonons could couple via ME interactions leading to a new
type of elementary excitation, electromagnon \cite{smolenskii82}.
Electromagnon has mixed characters of both magnons and infrared phonons.
It therefore opens a way to control
the magnetic excitations via electric fields \cite{mochizuki10a},
which may have important applications for
information processes.
However, despite electromagnon has been proposed theoretically almost four
decades ago, it has been observed experimentally only very recently
\cite{pimenov06}.
Till today, our knowledge about electromagnon is still very limited.

In the past few years, electromagnon has been observed in various
magnetoferroelectrics, mostly in the family of
RMnO$_3$\cite{pimenov06,aguilar07} and RMn$_2$O$_5$ \cite{sushkov07}
(R =Tb, Dy, Gd, Eu, Y) compounds.
So far, most theoretical works on electromagnon were
concentrated on the RMnO$_3$ family\cite{katsura07,aguilar09,mochizuki10},
because they have relatively simpler magnetic structures, containing only
Mn$^{3+}$ ions.
The study of electromagnon in RMn$_2$O$_5$ is still very few both
experimentally\cite{sushkov07,kim10} and theoretically \cite{sushkov08}, due to
the complexity of the magnetic structures in these compounds.
A simulation of the electromagnon in RMn$_2$O$_5$ at atomistic level
is still lack.
It has been shown experimentally that the dielectric step at the commensurate (CM) to
incommensurate (ICM) magnetic phase
transition in RMn$_2$O$_5$ is caused by the electromagnon \cite{sushkov07}.
However, it is still unclear how the electromagnon is related to the magnetodielectric
effect, which also happens at the same temperature
and is one of the most prominent effects in these materials \cite{hur04b}.

In this work, we carry out combined molecular-spin dynamics(MSD) simulations
to study electromagnons in RMn$_2$O$_5$.
We reproduce the most salient features of the electromagnon in these compounds
and find that the electromagnons frequencies
are very sensitive to the magnetic wave vector along the
$a$ direction.
We further investigate the behavior of
electromagnon under magnetic fields, and find that
although the mode frequency change
significant under magnetic field, the static dielectric constant
electromagnon does not change much in the magnetic field in our model.

%
To study the electromagnon in RMn$_2$O$_5$, we use the effective Hamiltonian derived in Ref.
\cite{wang08, cao09},
\begin{widetext}
\begin{equation}
H = \sum_{k} {1\over 2} m \omega^2 u_{k}^{2}-\sum_{ij \in
{J_{\alpha}} }J_{\alpha}(0) {\bf S}_i\cdot {\bf S}_j -\sum _{ij\in
{J_3}}\sum_{k} J_{3}' \, u_{k} {\bf S}_i\cdot {\bf S}_j
+ \sum_{kl}\xi_{kl} u_{k}\cdot u_{l}-\sum_{i}(K_{i}\cdot{\bf
S}_i)^2+\sum_{i}D_{i}(S_{i}^{c})^2-g\sum_i{\bf H}\cdot{\bf S}_i.
\label{eq:eff_h3}
\end{equation}
\end{widetext}
where $J_3$ is the Mn$^{4+}$- Mn$^{3+}$ superexchange interaction
through pyramidal base corners, and $J_4$ the superexchange
interaction through the pyramidal apex \cite{chapon04}. Mn$^{3+}$
couples to Mn$^{4+}$ either antiferromagnetically via $J_4$ along
$a$ axis or with alternating sign via $J_3$ along $b$ axis, whereas
Mn$^{3+}$ ions in two connected pyramids couple each other
antiferromagnetically through $J_5$. $J_1$, $J_2$ couple Mn$^{4+}$
ions along the $c$ axis. $u$ is an artificial optical phonon to
describe the electric polarization \cite{wang08,cao09}. 
In  RMn$_2$O$_5$, the infrared modes that couple to the
electromagnon are the same modes that causes the marcoscopic polarization. 
This is very different from those in  RMnO$_3$ compounds.
One can find more details of the model in Ref. \cite{cao09}. 
The major difference between the current model
and previous ones\cite{wang08,cao09} is that we include here the single ion easy
axis anisotropy $K_3$ for Mn$^{3+}$ ions and the easy plane anisotropies $D_i$
for Mn$^{3+}$ and Mn$^{4+}$ ions to describe the magnetic anisotropy in these
compounds. {\bf H} is the applied magnetic field and $g$ is
gyromagnetic ratio.

We use the model parameters obtained in Ref. \cite{cao09} until otherwise
noticed.
Since accurate
magnetic anisotropy energies are still very hard to obtain from the current
first-principles methods, especially for complex materials like RMn$_2$O$_5$,
we use empirical anisotropy parameters to fit the experimental
magnetization and optical spectra.

We carry out MSD simulations of the Hamiltonian
by solving the coupled equations of motion of spin and
phonon using forth-order Runge-Kutta method,
\begin{eqnarray}
\hbar\dot{{\bf S}_i} & = & -{\bf S}_i\times {\bf H}^{eff}_i \label{eq:EOMS}\\
\ddot{u}_k &= &{\bf F}_k/m \label{eq:EOM}
\end{eqnarray}
where ${\bf H}^{eff}_i=-\partial H /\partial {\bf S}_i $ is the
effective magnetic field acting on the $i$th Mn spin ${\bf S}_i$ and
${\bf F}_k=-\partial H /\partial u_k$ is the force acting on the
$k$th local phonon mode. Since Mn$^{3+}$ and Mn$^{4+}$ have 4 and 3
unpaired local $d$ electron respectively, we set the norm of
Mn$_3^{+}$ spin vector $|{\bf S}_{3}|=\sqrt{2(2+1)}$ and the norm of
Mn$_4^{+}$ spin vector $|{\bf S}_{4}|=\sqrt{1.5(1.5+1)}$.
Unlike previous methods \cite{mochizuki10},
our method takes the phonon degree of freedom into consideration explicitly.

We start the MSD simulation from a pool of equilibrium configurations
sampled from MC simulation at a given temperature $T$. The system is
further relaxed for sufficient time in the MSD simulations. 
Simulations with different
initial configurations are averaged to realize a canonical assemble
at fixed $T$ \cite{landau99}. The optical spectrum
Im$\chi(\omega)$ is calculated from the Fourier-Laplace transform of
correlation function $\Phi(t)$ of dipole moment P(t) with
$\Phi(t)=\langle P(t)P(0)\rangle-\langle P(t)\rangle \langle
P(0)\rangle$, i.e.,
\begin{equation}
Im \chi(\omega)=\frac{\omega Re{\mathcal{L}[\Phi]}}{K_{B}T}
\end{equation}
where $P(t)=\sum_{k}u_k(t)$ and $\langle \cdots \rangle$ represents
the ensemble average. The static susceptibility $\chi(0)$ can
be calculated as,
\begin{equation}
\chi(0)=\frac{\Phi(0)}{K_{B}T} \, ,
\end{equation}
where $K_B$ is the Boltzman constant.
To extract the detailed parameters of the spectral weight,
we fit $\chi(\omega)$ to a lorentz model,
\begin{eqnarray}
\chi(\omega)=\sum_j{\frac{S_j}{\omega_j^2-\omega^2-2i\gamma_{j}\omega}}
\label{eq:lorentz}
\end{eqnarray}
where $j$ enumerates the vibrations,
$\omega_j$ is the resonance frequency, and $\gamma_j$ is the damping
rate. To avoid numerical errors, we 
fit directly the correlation function $\Phi(t)$, 
which can be analytically Fourier transformed to lorentz model in
Eq.~\ref{eq:lorentz}. 
The static electric susceptibility is the sum of all oscillators strengths,
$\chi(0)=\sum_j\frac{S_j}{\omega^2_j}$. According to the
sum rule, the total spectral weight $S=\sum_jS_j$ should
be conserved.

\begin{figure}
\centering
\includegraphics[width=2.5in]{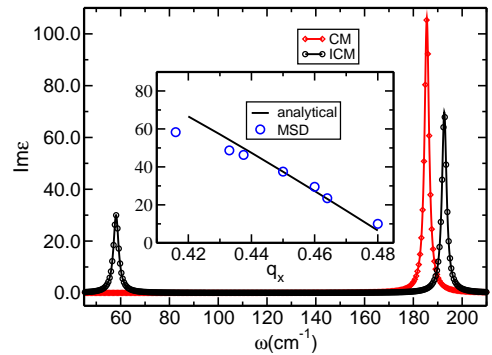}
\caption {(Color online). The imagery part of the dielectric response
Im$\chi(\omega)$ calculated using the $12 \times 12 \times 12$ lattice in the CM
magnetic phase (red) and the ICM magnetic phase (black). 
Inset: The electromagnon frequency as function 
of the magnetic wave vector along $a$ direction $q_x$. 
Circles represents the electromagnon frequency calculated by
MSD, whereas the black line is calculated from $J_{5}\sin(2\pi
q_x)- 14.63$ cm$^{-1}$. }
\label{fig:sus_CM_ICM}
\end{figure}

We first compare the MSD results to previous MC simulations \cite{cao09}, 
by carrying out the MSD simulation in
a 12$\times$12$\times$12 lattice at various
temperatures, without including the magnetic anisotropy.
The imaginary part of the dielectric response Im$\chi(\omega)$ of
the CM phase at 20K and ICM phase at 3K is shown in 
Fig.~\ref{fig:sus_CM_ICM}. We can see that, in the CM phase, there is only
one peak centered at about 185 cm$^{-1}$, corresponding to the
bare phonon frequency in our model. While at ICM phase, there are
two peaks: one is the phonon peak at about 193 cm$^{-1}$ which is
hardened compared to the bare phonon frequency due to strong
spin-phonon coupling \cite{footnote1}, 
and the new
peak appear at about 58 cm$^{-1}$ is the electromagnon.

\begin{figure}
\centering
\includegraphics[width=2.5in]{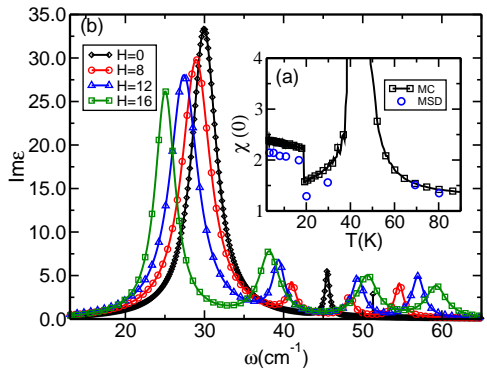}
\caption {(Color online) (a) The static dielectric susceptibility $\chi$(0) as
  functions of temperature calculated using MC and MSD.
(a) The imaginary part of the dielectric response Im $\varepsilon (\omega)$
of RMn$_2$O$_5$ under different magnetic fields.
} \label{fig:chi_omega}
\end{figure}

The calculated $\chi(0)$ is shown in
Fig. \ref{fig:chi_omega}(a)  at different temperatures compared with
those obtained from MC simulations \cite{cao09}. We see that
$\chi(0)$ calculated from both MC and MSD shows a step at the
CM to ICM transition.
The difference between $\chi(0)$
calculated with MC an MSD comes from the long-time relaxation effects
that are not included in the MSD simulation.  The results are in
remarkable agreement with the experimental results of
Ref. \onlinecite{sushkov07}. Although electromagnons have been
successfully reproduced using our model,
the frequency of the electromagnon mode
is much higher than the experimental values for
TbMn$_2$O$_5$ and YMn$_2$O$_5$\cite{sushkov07}.
Further examinations show that the
electromagnon frequencies are very sensitive to the $x$ component
of wave vector $q_x$. In the ICM phase, the spins connected by the
strongest interaction $J_4$ are almost parallel to each other\cite{cao09}, and the spiral
along $a$ direction comes from the spins connected by $J_5$, 
with an angle of 2$\pi q_x$. Since the electromagnon
excitation is at Brillouin zone center, 
$J_4$ makes no contribution to
the electromagnon as can be easily seen from Eq. \ref{eq:EOMS},
whereas the second largest exchange interaction $J_5$
dominates the interaction.
We therefore expect the energy of the electromagnon $\sim
J_{5}\sin(2\pi q_x)$. To confirm this result, we simulate the electromagnon
on a larger 50$\times$6$\times$6 lattice and adjust the parameter slightly
to obtain different $q_x$ of the ICM state.
The calculated lowest electromagnon mode frequencies at different $q_x$ are shown in the insert of
Fig. \ref{fig:sus_CM_ICM}.
As we see that the electromagnon frequencies decrease
significantly as $q_x$ approaching $\pi/2$, showing nice agreement with
$J_{5}\sin(2\pi q_x)-14.63$cm$^{-1}$.
At $q_x$=0.48, the electromagnon
frequency decreases to about 10 cm$^{-1}$, which is in good
agreement with experiments \cite{sushkov07}. 
Most of the RMn$_2$O$_5$ (R=Tb, Ho, ...) compounds have $q_x \sim$ 0.48,
therefore the electromagnon frequencies in these materials should be very similar.
However, we expect that the electromagnons in TmMn$_2$O$_5$\cite{kobayashi05}, with $q_x$ $\sim$ 0.467,
should have much higher frequencies. Experimental confirm of this prediction
is called for.

\begin{figure}
\centering
\includegraphics[width=2.6in]{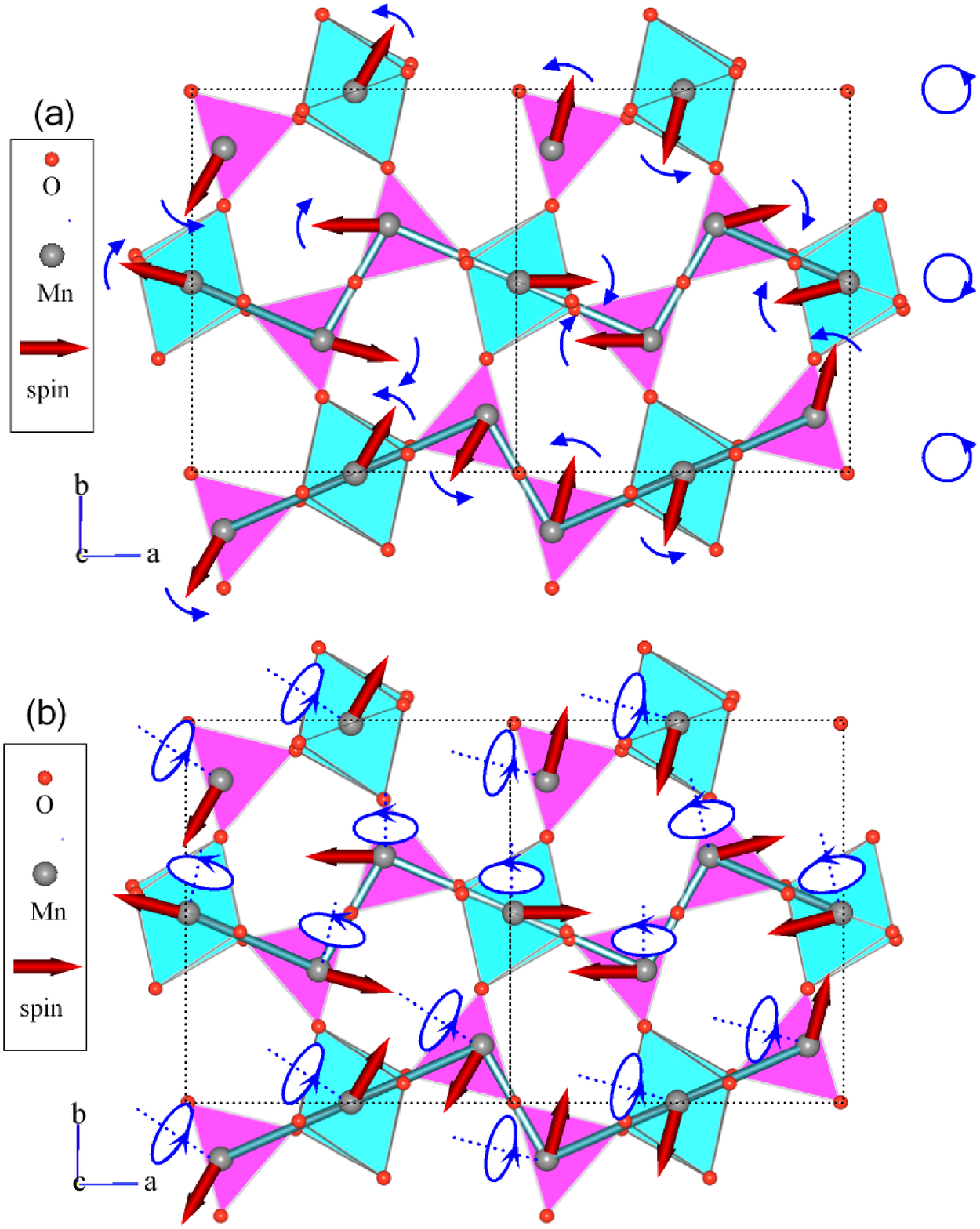}
\caption {(Color online) The cartoons of two identified electromagnon modes. 
The blue arrows denote rotation direction.
 (a) The ``phason'' mode, where spins in the same chain rotate in
  the same direction as a unit. 
(b) The highest frequency electromagnon mode appear under magnetic field H
  $\parallel a$. The blue arrows denote rotation directions.
} 
\label{fig:emnon}
\end{figure}

We now look into more details of the electromagonons in the ICM phase. To compare better
with the experiments,  we slightly modify our model
parameters to $J_3$=1.5 meV and $J_3^\prime$=0.3 meV to get $q_x$=0.46
in the 50$\times$6$\times$6 lattice \cite{footnote2}.
Though the frequencies are a little higher than the experimental data, it
does not change the basic physics of our model.
To see how the magnetic anisotropy affect the electromagnon,
we also include the ionic anisotropy energies in the simulation. 
We take $D_3=0.09$ meV, $D_4=0.06$ meV and three typical in-plane easy axis anisotropy energies 
$K_3$=(0.19, $\pm$ 0.21, 0), (0.278, $\pm$0.082, 0),
(0.08 $\pm$0.28, 0), 
corresponding to the case of Dy, Tb and Ho respectively. 
We first show the results using $K_3$=(0.19, $\pm$ 0.21, 0).

The imaginary part of the dielectric response
Im $\varepsilon(\omega)$ at 3K are presented in
Fig.~\ref{fig:chi_omega}.
After adding the magnetic anisotropy, 
the original electromagnon peak splits into three modes ,
centered at 30 cm$^{-1}$,
46 cm$^{-1}$, 51 cm$^{-1}$, respectively, consistent with recent
experiments \cite{sushkov07,kim10}. 
To extract the electromagnon modes, a short
electrical pulse $E\parallel b$ at $t=0$ is applied at the
equilibrium ground states.
The time evolution of each spin is
calculated and stored. The electromagnon normal modes are
reconstructed from the Fourier transform of each spin motion. From
the calculated mode profiles, we confirm that the strongest and
lowest frequency peak is the optical counterpart of ``phason'' with
wave vector {\bf Q}=0. The pattern of the mode is shown in Fig.\ref{fig:emnon}(a), 
which corresponds to the spin rotation in the
spiral plane \cite{kim10}. In this mode, the spins of
neighboring AFM chains rotate in opposite directions, 
and couple to phonon modes 
through the term $J_{3}' \,u{\bf S}_i\cdot {\bf S}_j\,$ forming
an electromagnon. We do not obtain a clear profile of
the other two modes, because they are relatively weak and
mixed with other modes.

Magnetodielectric effect is one of the most prominent effects in RMn$_2$O$_5$.
To clarify the relation between the electromagnon and magnetodielectric
effect, it is important to study the electromagnon in the magnetic field.
Figure~\ref{fig:chi_omega} depicts the imaginary part of dielectric response
for $H$=8, 12, 16 $T$ $\parallel a$ direction.
Interestingly, there is a new electromagnon peak emerges in the magnetic field
above the three electromagnon modes of the zero magnetic field.
The new mode can be characterized
as spins rotating around an axis perpendicular to itself in the $ab$
plane, as shown in Fig.\ref{fig:emnon}(b). 
We find that the pattern of the lowest frequency mode does not change.
and the magnitude of the intermediate two modes are getting stronger in
magnetic field. The frequencies of
the two lowest electromagnon peaks shift down with the
increase of magnetic field and the frequency of the lowest mode shifts down by about
17\% under $H$=16$T$ compared to that of zero field.
In contrast, the two high frequency modes shift
toward even higher frequency with increasing of the magnetic field.
To identify the driving force for electromagnon frequency
shifts, we compare the magnetic structure with and without
magnetic field and find that they are almost the same. Therefore,
the frequency shifts mainly come from the dynamic contribution of
Zeeman term $-g\sum_i{\bf H}\cdot{\bf S}_i$.

In spite of the large frequency change in magnetic field,
the total static susceptibility calculated from MSD remain almost unchanged
compared with that at $H$=0.
The total spectral weight also conserves well as the magnetic field increases,
within numerical errors.
The most contribution to $\chi(0)$ always comes from the lowest
electromagnon. At zero magnetic field,
the intermediate two electromagnon modes
have negligible contribution to the overall
$\chi(0)$. Under magnetic
field, the lowest electromagnon transfers spectral
weight to higher ones, e.g. $50\%$ at H=16 $T$. Therefore although the frequency decreases
significantly, $\chi(0)$ calculated from MSD almost keep constant.

We also study the electromagnon in magnetic fields with
H$\parallel b$  and H $\parallel c$.
The results for H $\parallel b$ are similar
to those for H $\parallel a$, while H $\parallel c$  has weak influence on
electromagnon both in frequency and spectral weight.
Our simulations suggest that the magnetodielectric effects in TbMn$_2$O$_5$
are not caused by the electromagnon, but it does not rule out such possibility
for DyMn$_2$O$_5$, which has unique magnetic structure in the ICM phase,
with $q_x$=0.5. 
We also do the simulations using two other $K_3$ parameters, and obtained very
similar results. 

The behaviors of electromagnons in RMn$_2$O$_5$
are very different from those in DyMnO$_3$ where
the electromagnon shows a soft-mode behavior in magnetic field
$H\parallel a$ and lead to the increase of dielectric
constant\cite{shuvaev10} in the field.
This is probably because the magnetic structures of
DyMnO$_3$ are much more sensitive to the applied magnetic field.

To summarize, we
investigated the electromagnon in magnetoferroelectrics RMn$_2$O$_5$ using
combined molecular-spin dynamics simulations. We have identified the origin of
the electromagnons in these compounds, and reproduced
the most salient features of the electromagnon in these compounds.
We find that the electromagnon frequencies are very sensitive to the
magnetic wave vector along the $a$ direction. We further investigate the
electromagnon in magnetic field. Although the mode frequency change
significant under magnetic field, the static dielectric constant
does not change much in the magnetic field in our model.

LH acknowledges the support from the Chinese National
Fundamental Research Program 2011CB921200 and
National Natural Science Funds for Distinguished Young Scholars.


\begin{thebibliography}{24}
\bibitem[{foo({\natexlab{a}})}]{footnote1}
\bibinfo{note}{The phonon mode in our model is an artificial one, which should
  not be compared directly to the experiments.}
\bibitem[{foo({\natexlab{b}})}]{footnote2}
\bibinfo{note}{Because $q_x$=0.48 corresponds to electromagnons with much lower
  frequencies, which need significantly longer simulation time. To reduce the
  computational time, we use $q_x$=0.46 other than 0.48, which does not change
  the essential physics in the simulations.}
\end{thebibliography}

\end{document}